\documentclass[
    ,final            % use final for the camera ready runs
sort
  ]
  {aipproc}

\layoutstyle{6x9}

\usepackage{graphicx}
\usepackage{graphics}
\usepackage{txfonts}
\usepackage{epsfig}
\usepackage{natbib}
\usepackage{times}
\usepackage{amssymb}

\newcommand\ergcms{erg\,cm$^{-2}$\,s$^{-1}$}

\newcommand\cmsq{cm$^{-2}$}

\newcommand\suz{{\it{Suzaku}}}

\newcommand\swift{{\it{Swift}}}

\newcommand\xmm{{\it{XMM-Newton}}}

\newcommand\nh{$N_\mathrm{H}$}

\begin{document}

\title{Supergiant X-ray binaries observed by \emph{Suzaku}}

\classification{98.70.Qy, 97.80.Jp, 97.60.Jd, 97.20.Pm, 97.10.Gz\vspace{-5mm}}
\keywords      {accretion, accretion disks ; gamma-rays: general ; stars: neutron ; supergiants ; X-rays: binaries ; X-rays: individual (IGR~J16207$-$5129; IGR~J17391$-$3021 = XTE~J1739$-$302)}

\author{A. Bodaghee}{
  address={Space Sciences Laboratory, University of California, Berkeley, USA},
  email={bodaghee@ssl.berkeley.edu},
}

\author{J. A. Tomsick}{
  address={Space Sciences Laboratory, University of California, Berkeley, USA},
}

\author{J. Rodriguez}{
  address={Lab. AIM, CEA/IRFU - UPD -- CNRS/INSU, CEA DSM/IRFU/SAp, Centre de Saclay, France} 
}

\author{S. Chaty}{
  address={Lab. AIM, CEA/IRFU - UPD -- CNRS/INSU, CEA DSM/IRFU/SAp, Centre de Saclay, France} 
}

\author{K. Pottschmidt}{
  address={CRESST/UMBC, NASA Goddard Space Flight Center, USA}
}

\author{R. Walter}{
  address={ISDC, Observatoire de l'Universit\'e de Gen\`eve, Switzerland}
}

\author{P. Romano}{
  address={INAF, Istituto di Astrofisica Spaziale e Fisica Cosmica, Palermo, Italy}
}

\begin{abstract}
\suz\ observations are presented for the high-mass X-ray binaries IGR~J16207$-$5129 and IGR~J17391$-$3021. For IGR~J16207$-$5129, we provide the first X-ray broadband (0.5--60 keV) spectrum from which we confirm a large intrinsic column density (\nh\ $= 1.6\times10^{23}$\,\cmsq), and we constrain the cutoff energy for the first time ($E_{\mathrm{cut}} = 19$\,keV). A prolonged ($>$ 30 ks) attenuation of the X-ray flux was observed which we tentatively attribute to an eclipse of the probable neutron star by its massive companion, in a binary system with an orbital period between 4 and 9 days, and inclination angles $>$ 50 degrees. For IGR~J17391$-$3021, we witnessed a transition from quiescence to a low-activity phase punctuated by weak flares whose peak luminosities in the 0.5--10\,keV band are only a factor of 5 times that of the pre-flare emission. These micro flares are accompanied by an increase in \nh\ which suggests the accretion of obscuring clumps of wind. We now recognize that these low-activity epochs constitute the most common emission phase for this system, and perhaps in other supergiant fast X-ray transients (SFXTs) as well. We close with an overview of our upcoming program in which \suz\ will provide the first ever observation of an SFXT (IGR~J16479$-$4514) during a binary orbit enabling us to probe the accretion wind at every phase.
\end{abstract}

\maketitle

\section{Introduction \& Observations}
\vspace{-2mm}

High-mass X-ray binaries (HMXBs) are compact objects (usually a neutron star but sometimes a black hole) which accrete from a massive donor star. An HMXB can be further classified based on whether this donor is a B-emission line star (BEXBs) or a supergiant OB star (SGXBs). Recently, another subclass of HMXB has been recognized wherein the donor star is clearly a supergiant OB star as in SGXBs, but the X-ray emission shows a large degree of variability as in BEXBs. These systems are called supergiant fast X-ray transients (SFXTs). 

\suz, with its high sensitivity over a broad X-ray bandpass, can help resolve several outstanding issues concerning the accretion physics and nature of HMXBs. We point the reader to \suz\ observations which: confirmed the presence of clumpy winds in the SFXT called IGR~J17544$-$2619 \citep{Ram09}; provided the most accurate broadband X-ray spectrum of the highly-obscured IGR~J16318$-$4848 \citep{Bar09}; and which found a low-level of activity in the SFXT IGR~J08408$-$4503 even during quiescence \citep{Sid10}. In 2008, \suz\ observed the SGXB IGR~J16207$-$5129 and the SFXT IGR~J17391$-$3021 (= XTE~J1739$-$302) for 33 and 37 ks of effective exposure time, respectively. The analysis of these sources is detailed in \citet{Bod10,Bod11} and is summarized below.

\section{IGR~J16207$-$5129}
\vspace{-3mm}

%__________________________________________________________________16207
\begin{figure*}[!b] 
\includegraphics[height=0.3\textheight]{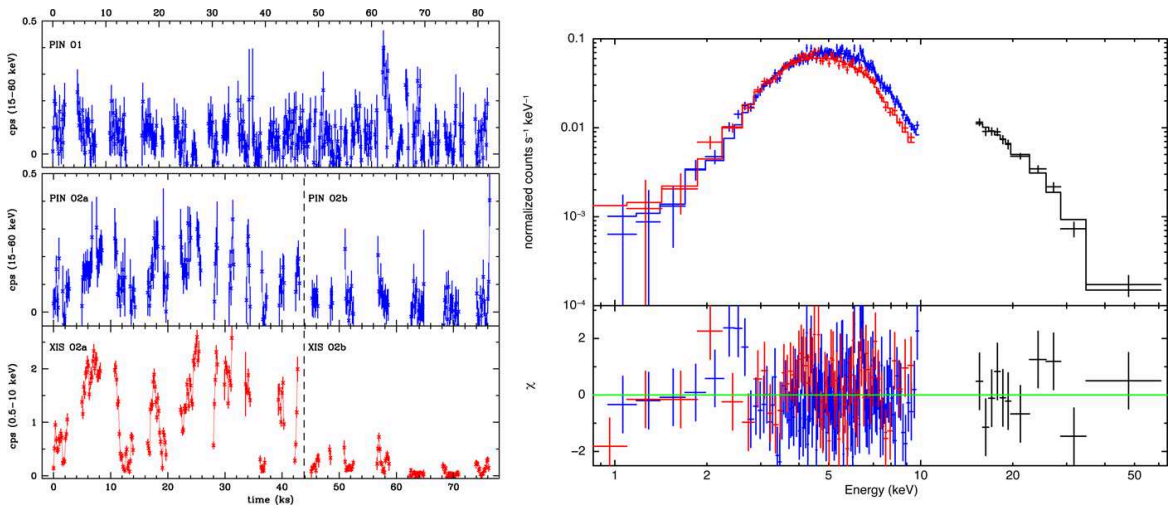}
\caption{\emph{Left}: Background-subtracted light curve of IGR~J16207$-$5129 from XIS (red: 0.5--10 keV) and HXD (blue: 15--60 keV). The upper panel displays the first observation (O1) while the lower panel shows the second observation (O2) which includes simultaneous data from XIS. Each bin collects 160 s worth of data. The dashed line represents MJD 54527.382. This corresponds to the onset of an unusually long period ($>$ 30 ks) of suppressed flux which could represent an eclipse of the compact object by its supergiant stellar companion. \emph{Right}: Spectrum of IGR~J16207$-$5129 corrected for the background and fit with an absorbed power law with an exponential cutoff whose parameters are listed in Table\,1. Each bin collects at least 150 photon counts from XIS-1 (red), XIS-0 paired with XIS-3 (blue), and PIN (black).}
\label{fig_16207}
\end{figure*}

Figure \ref{fig_16207} shows that the emission from IGR~J16207$-$5129 varies within 1--2 orders of magnitude on ks timescales, which is typical of SGXBs. Intriguingly, X-rays from the source, and the variability associated with that emission, are significantly diminished during the last $\sim$30 ks of the observation. If the accretor were to enter an ``off'' state, this would last a few 100 s, or at least an order of magnitude shorter than the period of inactivity that we observed. The observed attenuation of the X-ray flux can not be explained by an increase in the column density (i.e., occulting wind clumps) since the measured \nh\ does not change in a significant way between O2a and O2b (Table \ref{tab_spec}). An occultation of the neutron star would imply unrealistic column densities at other orbital phases. A remaining viable explanation is an eclipse of the compact object by its companion, but the data do not allow us to confirm or reject this hypothesis. If the eclipse is real, it lasts at least 30 ks (0.35 d) setting a lower limit on the orbital period at 0.7 d. With this constraint, and assuming typical stellar parameters, a model for an eclipsing binary \citep{Rap83} yields orbital periods between 4 and 9 d with inclination angles $>$ 50$^{\circ}$. We note that whatever mechanism is involved in driving the sporadic emission in SFXTs (e.g., the 33-ks dormant period of IGR~J17391$-$3021 in Fig. \ref{fig_17391}) could be responsible for the prolonged attenuation seen in IGR~J16207$-$5129. The spectrum of IGR~J16207$-$5129 (Fig. \ref{fig_16207}) features a large absorbing column (\nh\ $> 10^{23}$\,\cmsq) and a possible iron line, both of which suggest the presence of matter around the X-ray source. A cutoff near 20 keV favors a neutron star as the compact object, but coherent pulsations or cyclotron absorption lines that would confirm the presence of a neutron star are not detected.

\section{IGR~J17391$-$3021}
\vspace{-3mm}

%__________________________________________________________________17391
\begin{figure*}[!b] 
\includegraphics[height=0.3\textheight]{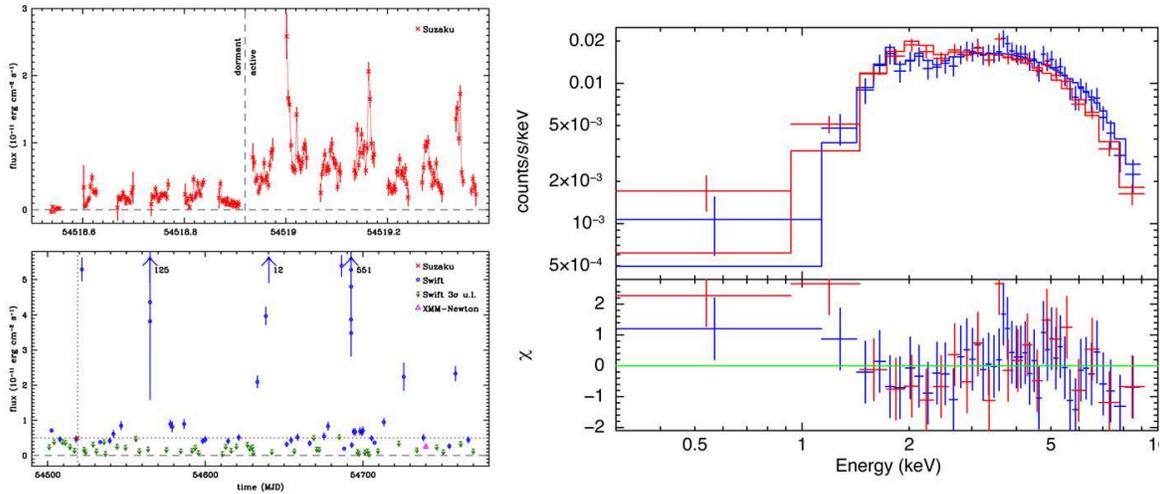}
\caption{\emph{Left}: Light curve of IGR~J17391$-$3021 from XIS, \swift-XRT, and \xmm. The top panel presents the background-subtracted light curve from our \suz\ observation (red crosses, 240-s binning). In the bottom panel, the source light curve from the \swift\ monitoring campaign \citep{Rom11} is shown (blue circles, $\sim$1 ks of exposure time per data point) along with their 3-$\sigma$ upper limits for non-detections (green downward arrows). The blue upwards arrows designate the maximum flux of large outbursts detected by \swift\ that are situated beyond the scale of the graph. The average flux from our 37-ks \suz\ observation is plotted as a single red cross at the intersection of the dotted lines. The magenta triangle at MJD 54740 corresponds to the average flux in a 31-ks observation with \xmm\ \citep{Boz10}. Fluxes are given in the 0.5--10 keV energy band as observed in units of $10^{-11}$ \ergcms. \emph{Right}: Background-corrected spectrum of IGR~J17391$-$3021 fit with an absorbed power law. The data represent photon counts from XIS-1 (red), and XIS-0 combined with XIS-3 (blue). Each bin collects a minimum of 150 counts.}
\label{fig_17391}
\end{figure*}

During the initial 33 ks of the observation of IGR~J17391$-$3021, the source is in an extremely low-activity phase that is at or near quiescence (Fig. \ref{fig_17391}). The source then enters a period of enhanced activity in which the luminosity is only a factor of 5 that of the ``quiescent'' emission. \citet{Boz10} noticed similar behavior in a recent \xmm\ observation of this source. Observations prior to and after MJD 54518.92 are referred to henceforth as ``dormant'' and ``active'' epochs, respectively. The active state features 3 quasi-periodic flares separated by $\sim$15 ks intervals. The peak fluxes are at the level of the faint detections (and a few upper limits) from \swift\ monitoring \citep{Rom11}. Epochs of enhanced activity just above quiescence, but well below the bright flaring episodes typical of this class, represent over 60\% of all observations, so they are the most common emission state. Given more exposure time, many \swift\ upper limits would be detections suggesting that the duty cycle is even higher. Based on the source ephemeris \citep{Dra10}, which places the \xmm\ and \suz\ observations at different phases, such low-activity states are not confined to a specific part of the orbit. Table \ref{tab_spec} shows that when IGR~J17391$-$3021 is in the active state, the \nh\ is at least twice as high as it is in the dormant phase. This is unlike what was seen by \citet{Boz10} who found that the \nh\ remained steady (within the statistical errors) during weak flares while $\Gamma$ varied by around 50\%. Differences in the spectral properties of the flares caught by \suz\ and \xmm\ could be due to unequal geometric configurations of the system between the observations. The \nh\ during activity is 2--4 times the interstellar value which suggests that the source is not strongly absorbed (intrinsically) except when the compact object accretes a clump passing along our line of sight \citep[see also][]{Ram09}.

\vspace{-3mm}
\section{Conclusions \& Perspectives}
\vspace{-3mm}

The putative eclipse in the persistently-emitting SGXB IGR~J16207$-$5129 awaits confirmation \citep{Bod10}. Micro-flares from IGR~J17391$-$3021 hold valuable clues to the accretion processes of these intriguing transients \citep{Bod11}. The demarcation between SGXBs and SFXTs is not as clear as originally believed, and a few systems such as IGR~J16479$-$4514 might represent an intermediate state. In this respect, our upcoming 150-ks long \suz\ observation of IGR~J16479$-$4514 at various phases of its orbit will help us understand the extent to which perceived emissivity differences between these populations stem from their unequal wind and orbital characteristics. 

\begin{table}
\begin{tabular}{r r r r r r}
\hline
     \tablehead{1}{r}{b}{source}
  & \tablehead{1}{r}{b}{epoch}
  & \tablehead{1}{r}{b}{\nh\ \\$10^{22}$\,cm$^{-2}$}
  & \tablehead{1}{r}{b}{$\Gamma$}
  & \tablehead{1}{r}{b}{$L$ \\$10^{33}$\,erg\,s$^{-1}$	}
  & \tablehead{1}{r}{b}{$\chi_{\nu}^{2}/$dof}   \\
\hline
IGR~J16207$-$5129	& total 		& 16$\pm$1			& 0.9$\pm$0.2 		& 130				&  1.05/199		 	\\	
				& O2a  		& 19$\pm$1			& 1.3$\pm$0.1 		& 240				&  0.91/182		 	\\	
				& O2b 	 	& 19$\pm$5			& 1.5$\pm$0.4 		& 56					&  0.51/64		 	\\
\hline
IGR~J17391$-$3021	& total 		& 3.6$\pm$0.4			& 1.4$\pm$0.1 		& 4.8					&  0.82/66		 	\\	
				& dormant 	& 1.0$\pm$0.6			& 1.0$\pm$0.3 		& 1.3					&  0.68/8		 	\\	
				& active 		& 4.1$\pm$0.5			& 1.5$\pm$0.1 		& 7.4					&  0.73/54		 	\\	
\hline
\end{tabular}
\caption{Parameters from absorbed power laws fit to the \suz\ spectra of IGR~J16207$-$5129 (0.5--60\,keV) and IGR~J17391$-$3021 (0.5--10\,keV) for various epochs. The luminosity in the 0.5--10\,keV band is corrected for absorption. Errors represent 90\% confidence.}
\label{tab_spec}
\end{table}

\bibliographystyle{aipproc}  
\vspace{-1mm}

\bibliography{Bodaghee_Arash}

\end{document}